\documentclass[print, twocolumn]{revtex4}
\usepackage{amsmath}
\usepackage{txfonts}
\usepackage{graphicx}
\usepackage{bm}
\usepackage{epsfig}
\usepackage{setspace}
\begin{document}
\draft
\bibliographystyle{revtex4}

\titlepage
\title{Dynamics of quantum correlations for two-qubit coupled to a spin chain with Dzyaloshinskii-Moriya interaction}
\author{Yi-Ying Yan$^{1}$}\author{Li-Guo Qin$^{1}$}\author{Li-Jun Tian\footnote{Email: tianlijun@staff.shu.edu.cn.}$^{1,2}$}
\affiliation {$^{1}$Department of Physics, Shanghai University, Shanghai,
200444, China}\affiliation{$^{2}$Shanghai Key Lab for Astrophysics, Shanghai,
200234, China}

\begin{abstract}
We study the dynamics of quantum discord and entanglement for two spin qubits coupled to a
spin chain with Dzyaloshinsky-Moriya (DM) interaction. We
numerically and analytically investigate the time evolution of quantum discord and entanglement for two-qubit initially prepared in a class of $X-$structure state. In the case of evolution from a pure state, quantum correlations decay to zero in a very short time at the critical point of the environment. In the case of evolution from a mixed state, It is found that quantum discord may get maximized due to the quantum critical behavior of the environment while entanglement vanishes under the same condition. Moreover, we observed sudden transition between classical and quantum decoherence when single qubit interacts with the environment. The effects of DM interaction on quantum correlations are also considered and revealed in the two cases. It can enhance the decay of quantum correlations and its effect on quantum correlations can be strengthened by anisotropy parameter.
\end{abstract}

\maketitle

\section{Introduction}
Entanglement is thought to be the fundamental resource for quantum
computation and communication \cite{Nielsen1}. It is well known as a
kind of quantum correlation, which is original from the
superposition principle of quantum mechanics. Recently, quantum
discord is realized as a different kind of quantum correlation other
than entanglement, it is arising from the difference between
quantum mutual information and maximum of quantum conditional
entropy \cite{Zurek1, Vedral1}. This nonentanglement correlation may be
applied to the quantum computation, which is based on deterministic quantum computation with one quantum
bit (DQC$1$) protocol \cite{Datta1, Lanyon1}. Moreover, A. Brodutch \textit{et al}. \cite{Brodutch1} have shown that
the changes in quantum discord is pointed to be an indicator of failure of a
local operations and classical communications (LOCC) implementation of the quantum gates. Both entanglement and quantum
discord characterize the same quantumness features of quantum
system in the pure state, namely, quantum discord reduces to be entropy of entanglement. However, they become two different
measures for quantum system in the mixed state. For instance, quantum discord may be
nonzero in the disentangled state \cite{Luo1, Ali1}.

The properties of quantum discord are intensively investigated due
to its potential application. It is realized that quantum discord
may be generated in the Heisenberg models at finite temperature
\cite{Ciliberti1, Maziero1, Hassan1, Pal1, Werlang2}. Quantum discord behaves in a very different way
from that of entanglement by changing the parameters. For instance, the situation where
quantum discord increases with temperature while entanglement
decreases is revealed \cite{Werlang2}. Entanglement is also known as
an indicator to signal a quantum phase transition (QPT) \cite{Vidal2}, so does the quantum discord. Sarandy has analyzed the quantum discord in the critical systems, such as XXZ and transverse field Ising models, and shown that quantum discord exhibit signature of the QPT \cite{Sarandy1}. Furthermore, Werlang \textit{et al}. have show that quantum discord, in contrast to entanglement, spotlight the critical points associated with QPT for Heisengber model even at finite temperature \cite{Werlang3}. More recently, It is found that quantum discord detect quantum critical points associated with first- and higher-order QPTs caused by field and three-spin interactions at finite temperature \cite{Li2}.

Decoherence of the quantum system due to interacting with its environment is the main
obstacle to practice quantum computation tasks. Thus, it is necessary to reveal the
dynamical properties of quantum discord for designing the protocol to against decoherence.
The dynamics of quantum discord both in the Markovian and nonMarkovian environment are theoretically and experimentally
investigated \cite{Werlang1, Fanchini1, Wang1, Xu1}. The researchers have found the situation where quantum discord disappears
in the asymptotic time limit but entanglement undergoes a sudden death \cite{Werlang1}, which is found in the Morkovian environment.
In this sense, quantum discord is thought to be more robust than entanglement. Recently, there is an increasing investigation on decoherence due to spin environment \cite{Hutton1, Cucchietti1, Rossini1, Quan1, Yuan1, Cheng1, Liu1}, such as single qubit coupled to the environment and two qubits coupled to the environment. It is revealed that quantum coherence is dramatically destroyed in the critical point of the QPT from the external environment.

In this paper, we investigate the dynamical behavior of quantum
discord and entanglement for two-qubit coupled to a spin chain with DM
interaction. The DM interaction is arising from the spin-orbit coupling \cite{Moriya1}, which often appears in the models of low-dimensional magnetic materials. The dynamics of the system also depends on the initial state. We choose the two-qubit to be a class of $X-$structure state, which is general to contain Bell-diagonal state and Werner state, which are very important and famous in the quantum information theory. We start in section \ref{sec2} by introducing and diagonalizing the model, and we calculate the reduced density matrix of two-qubit. In section \ref{sec3}, we
analytically and numerically evaluate quantum discord and entanglement, and present the main results and discussion. The last section is devoted to the conclusions.

\section{Model and its Solution}\label{sec2}
We consider the environment to be an general $XY$ chain with $z$-component DM interaction.
The total Hamiltonian for two spin qubits transversely coupled to the chain is given by
\begin{equation}\label{e0}
H=H^{(\lambda)}_{E}+H_{I}
\end{equation}
with
\begin{eqnarray}\label{e1}
H^{(\lambda)}_{E}&=&-\sum^{N}_{l}\bigg(\frac{1+\gamma}{2}\sigma^{x}_{l}\sigma^{x}_{l+1}+\frac{1-\gamma}{2}\sigma^{y}_{l}\sigma^{y}_{l+1}+\lambda\sigma^{z}_{l}\bigg)  \nonumber\\
& &-\sum^{N}_{l}D(\sigma^{x}_{l}\sigma^{y}_{l+1}-\sigma^{y}_{l}\sigma^{x}_{l+1}), \\
H_{I}&=&-g(\frac{1+\delta}{2}\sigma^{z}_{A}+\frac{1-\delta}{2}\sigma^{z}_{B})\sum^{N}_{l}\sigma^{z}_{l},
\end{eqnarray}
where $H^{(\lambda)}_{E}$ is the self-Hamiltonian of the environment, $H_{I}$ describes
the interaction between two-qubit and environment. $\sigma^{\alpha}_{l} (\alpha=x, y, z)$ is the Pauli
matrix on $l-$th sites of the chain. $\sigma^{z}_{A(B)}$ describes the two-qubit. $\gamma$ measures anisotropy
of exchange interaction in the $XY$ plane, $\lambda$ is the strength of transverse magnetic field, $D$ is the strength
of $z-$component DM interaction. $g\frac{1+\delta}{2} (g\frac{1-\delta}{2})$ describes the coupling strength between qubit $A(B)$ and surrounding chain. The parameter $\delta$ controls anisotropy of coupling strength of qubit with its environment.
$N$ is the total sites of $XY$ chain. The periodic boundary conditions $\sigma^{\alpha}_{N+1}=\sigma^{\alpha}_{1}$ is assumed.
The eigenstates of the operator $(\frac{1+\delta}{2}\sigma^{z}_{A}+\frac{1-\delta}{2}\sigma^{z}_{B})$ are simply given by
\begin{eqnarray}\label{e2}
|\phi_{1}\rangle=|00\rangle, |\phi_{2}\rangle=|01\rangle, |\phi_{3}\rangle=|10\rangle,|\phi_{4}\rangle=|11\rangle,
\end{eqnarray}
where $|0\rangle$ and $|1\rangle$ denote spin up and down respectively.
In terms of these states, the Hamiltonian (\ref{e0}) can be rewritten as
\begin{equation}\label{e3}
H=\sum^{4}_{\mu=1}|\phi_{\mu}\rangle\langle\phi_{\mu}|\otimes H^{(\lambda_{\mu})}_{E},
\end{equation}
where the parameters $\lambda_{\mu}$ are
\begin{equation}
\lambda_{1(4)}=\lambda\pm g, \lambda_{2(3)}=\lambda\pm g\delta,
\end{equation}
and $H^{(\lambda_{\mu})}_{E}$ is obtained from $H^{(\lambda)}_{E}$ by replacing $\lambda$ with $\lambda_{\mu}$.
To investigate the dynamics of two-qubit, we need to compute the time evolution operator $U(t)=\exp(-iHt)$.
For the purpose, we map the projected Hamiltonian $H^{(\lambda_{\mu})}_{E}$ into a one-dimensional spinless fermion
system with creation and annihilation operators $c^{\dag}_{l}$ and $c_{l}$ via Jordan-Wigner transformation \cite{Sachdev1}
\begin{eqnarray}\label{e4}
\sigma^{x}_{l}&=&\prod_{s<l}(1-2c^{\dag}_{s}c_{s})(c_{l}+c^{\dag}_{l}), \nonumber \\
\sigma^{y}_{l}&=&-i\prod_{s<l}(1-2c^{\dag}_{s}c_{s})(c_{l}-c^{\dag}_{l}), \nonumber \\
\sigma^{z}_{l}&=&1-2c^{\dag}_{l}c_{l}.
\end{eqnarray}
Following by Fourier transforms of the fermionic operator given by $d_{k}=\frac{1}{\sqrt{N}}\sum^{N}_{l}c_{l}e^{-i2\pi lk/N}$ with $k=-M,\ldots,M$ and $M=(N-1)/2$, the Hamiltonian is transformed to momentum space, and then using Bogoliubov transformation
\begin{equation}\label{e5}
\eta_{k,\lambda_{\mu}}=\cos\frac{\theta^{(\lambda_{\mu})}_{k}}{2}d_{k}-i\sin \frac{\theta^{(\lambda_{\mu})}_{k}}{2}d^{\dag}_{-k},
\end{equation}
with $\theta^{(\lambda_{u})}_{k}=\arctan\bigg(\frac{\gamma\sin\frac{2\pi k}{N}}{\lambda_{\mu}-\cos\frac{2\pi k}{N}}\bigg)$, finally we get the diagonalized Hamiltonian
\begin{equation}
H^{(\lambda_{\mu})}_{E}=\sum_{k}\Lambda^{(\lambda_{\mu})}_{k}(\eta^{\dag}_{k,\lambda_{\mu}}\eta_{k,\lambda_{\mu}}-\frac{1}{2}),
\end{equation}
where the spectrum $\Lambda^{(\lambda_{\mu})}_{k}$ is $\Lambda^{(\lambda_{\mu})}_{k}=2(\varepsilon^{(\lambda_{\mu})}_{k}+2D\sin\frac{2\pi k}{N})$ with $\varepsilon^{(\lambda_{\mu})}_{k}=\sqrt{(\lambda_{\mu}-\cos\frac{2\pi k}{N})^{2}+\gamma^{2}\sin^{2}\frac{2\pi k}{N}}$.

Suppose the initial state of total system is disentangled with $\rho_{tot}(0)=\rho_{AB}(0)\otimes\rho_{E}(0)$,
where $\rho_{AB}(0)$ and $\rho_{E}(0)$ are the initial state of two-qubit system and environment, respectively.
$\rho_{E}(0)=|\psi_{E}(0)\rangle\langle\psi_{E}(0)|$ is assumed to be the groundstate of the environment.
The evolution of the total system is governed by $\rho_{tot}(t)=U(t)\rho_{tot}(0)U^{\dag}(t)$.
Then the reduced density matrix of two-qubit $AB$ is obtained by tracing out the environment
\begin{eqnarray}\label{e6}
\rho_{AB}(t)&=&Tr_{E}[\rho_{tot}(t)] \nonumber \\
&=&\sum^{4}_{\mu,\nu}F_{\mu\nu}(t)\langle \phi_{\mu}|\rho_{AB}(0)|\phi_{\nu}\rangle|\phi_{\mu}\rangle \langle\phi_{\nu}|
\end{eqnarray}
with $F_{\mu\nu}(t)=\langle\psi_{E}|U^{\dag(\lambda_{\nu})}_{E}(t)U^{(\lambda_{\mu})}_{E}(t)|\psi_{E}\rangle$, where $U^{(\lambda_{\mu})}_{E}(t)=\exp(-iH^{(\lambda_{\mu})}_{E}t)$ is the projected time evolution operator driven by $H^{(\lambda_{\mu})}_{E}$.

Now we suppose that the two-qubit are initially prepared in the $X-$structure states,
\begin{equation}\label{e7}
\rho_{AB}(0)=\frac{1}{4}\bigg(I_{AB}+\sum_{\alpha}c_{\alpha}\sigma_{A}^{\alpha}\otimes\sigma_{B}^{\alpha}\bigg)
\end{equation}
with $I_{AB}$ is the identity operator on two-qubit system, $\alpha=x,y,z$, and the parameters $c_{\alpha}$ are
choose to be real parameters that promise the $\rho_{AB}(0)$ be a legal quantum state.
This state is general to contain Bell-diagonal states and Werner states.
According to Eq.~(\ref{e6}) and (\ref{e7}), the reduced density matrix can be written in the
standard basis $\{|00\rangle,|01\rangle,|10\rangle,|11\rangle\}$
\begin{equation}\label{e8}
\rho_{AB}(t)=\frac{1}{4}\left(\begin{array}{cccc}
1+c_{z} & 0 & 0 & \Gamma\\
0 & 1-c_{z} & \Omega & 0\\
0 & \Omega^{*} & 1-c_{z} & 0\\
\Gamma^{*} & 0 & 0 & 1+c_{z}
\end{array}\right)
\end{equation}
with $\Gamma=(c_{x}-c_{y})F_{14}(t)$ and $\Omega=(c_{x}+c_{y})F_{23}(t)$. To evaluate the quantum discord in this
density matrix, it is necessary to compute the decoherence factor $F_{\mu\nu}(t)$. Let $|G\rangle_{\lambda}$ and $|G\rangle_{\lambda_{\mu}}$
denote the groundstates of the self-Hamiltonian $H^{(\lambda)}_{E}$ and the projected-Hamiltonian, respectively.
By using the transformation \cite{Yuan1,Quan1} $|G\rangle_{\lambda}=\prod^{M}_{k>0}(\cos\Theta^{(\lambda_{\mu})}_{k}+i\sin\Theta^{(\lambda_{\mu})}_{k}\eta^{\dag}_{k,\lambda_{\mu}}\eta^{\dag}_{-k,\lambda_{\mu}})
|G\rangle_{\lambda_{\mu}}$ with $\Theta^{(\lambda_{\mu})}=(\theta^{(\lambda_{\mu})}_{k}-\theta^{(\lambda)}_{k})/2$
and following a tedious calculation, we obtain the decoherence factor as \cite{Quan1}
\begin{eqnarray}\label{e9}
|F_{\mu\nu}(t)| & = & \prod_{k>0}^{M}[1-\sin^{2}(2\Theta_{k}^{(\lambda_{\mu})})\sin^{2}(\Lambda_{k}^{(\lambda_{\mu})}t)-\sin^{2}(2\Theta_{k}^{(\lambda_{\nu})}) \nonumber \\
 &  & \times \sin^{2}(\Lambda_{k}^{(\lambda_{\nu})}t)+2\sin(2\Theta_{k}^{(\lambda_{\mu})})\sin(2\Theta_{k}^{(\lambda_{\nu})}) \nonumber \\
 &  & \times \sin(\Lambda_{k}^{(\lambda_{\mu})}t)\sin(\Lambda_{k}^{(\lambda_{\nu})}t)\cos(\Lambda_{k}^{(\lambda_{\mu})}t-\Lambda_{k}^{(\lambda_{\nu})}t)-4\nonumber \\
 &  & \times \sin(2\Theta_{k}^{(\lambda_{\mu})})\sin(2\Theta_{k}^{(\lambda_{\nu})})\sin^{2}(\Theta_{k}^{(\lambda_{\mu})}-\Theta_{k}^{(\lambda_{\nu})}) \nonumber \\
 &  & \times \sin^{2}(\Lambda_{k}^{(\lambda_{\mu})}t)\sin^{2}(\Lambda_{k}^{(\lambda_{\nu})}t)]^{1/2}\equiv \prod^{M}_{k>0}F_{k}(t).
\end{eqnarray}
When $|F_{\mu\nu}(t)|\rightarrow1$, it reveals that the two-qubit is slightly
disturbed by environment. When $|F_{\mu\nu}(t)|\rightarrow0$, the two qubtis system undergoes
strong decoherence due to the environment. We present the detail results and discussion in the next section.

\begin{figure}
  \includegraphics[width=9cm]{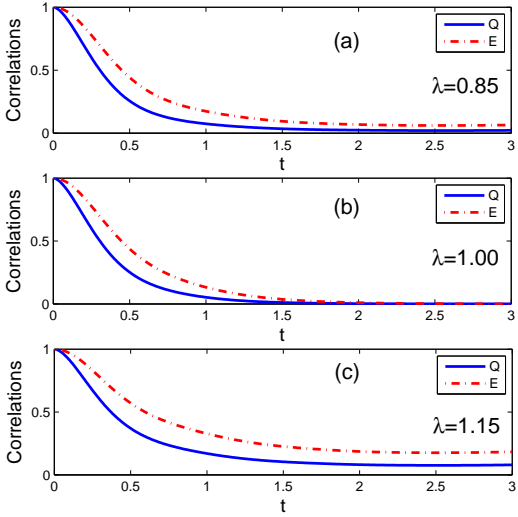}\\
  \caption{Quantum discord (solid line) and entanglement(dash-dot line) as a function of time $t$ for three different $\lambda$. Other parameters are set as $\gamma=1, g=0.05, \delta=0, D=0$, and $N=600$. (a) $\lambda=0.85$, (b) $\lambda=1.00$, (c) $\lambda=1.15$. }\label{FIG1}
\end{figure}

\begin{figure}
  \includegraphics[width=8cm]{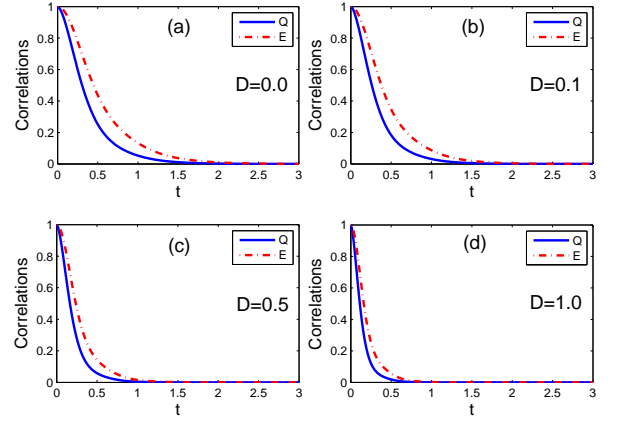}\\
  \caption{Quantum discord (solid line) and entanglement (dash-dot line) as a function of time $t$ under different $D$. Other parameters are set as $\gamma=1, \lambda=1, g=0.05, \delta=0$, and $N=600$, (a) $D=0.0$, (b) $D=0.1$, (c) $D=0.5$, (d) $D=1.0$.}\label{FIG2}
\end{figure}

\section{Correlations for the two-qubit}\label{sec3}
Quantum discord is proposed to measure quantumness correlation in the bipartite quantum system.
quantum discord for two-qubit state $\rho_{AB}$ is defined as \cite{Zurek1},
\begin{equation}
Q(\rho_{AB})=I(\rho_{AB})-C(\rho_{AB}),
\end{equation}
where $I(\rho_{AB})=S(\rho_{A})+S(\rho_{B})-S(\rho_{AB})$ with $S(\rho)=-Tr(\rho\log_{2}\rho)$ is the
quantum mutual information, which measures the total correlation in state $\rho_{AB}$. $\rho_{A(B)}$ is the reduced
density matrix of $\rho_{AB}$. $C(\rho_{AB})$ in the above definition is the maximum of quantum conditional entropy
by performing one side measurement \cite{Vedral1},
\begin{equation}
C(\rho_{AB})=\max_{\{\Pi_{B}^{i}\}}\bigg\{S(\rho_{A})-\sum_{i}p_{i}S(\rho_{A}^{(i)})\bigg\},
\end{equation}
where $\{\Pi^{i}_{B}\}$ denotes a set of von Neumann measurement on $B$, $\rho_{A}^{(i)}=Tr_{B}(\Pi_{B}^{i}\rho_{AB}\Pi_{B}^{i})/p_{i}$ is the state of $A$ after obtaining outcome $i$ on $B$, where $p_{i}= Tr_{AB}(\Pi_{B}^{i}\rho_{AB}\Pi_{B}^{i})$. $C(\rho_{AB})$ is suggested to measure classical correlation in the state $\rho_{AB}$.

To obtain the quantum discord of $\rho_{AB}(t)$, we need to calculate the quantum mutual information
and classical correlation. The quantum mutual information is directly calculated,
\begin{equation}
I[\rho_{AB}(t)]=2+\sum^{4}_{n=1}\omega_{n}\log_{2}\omega_{n},
\end{equation}
with $\omega_{1}=\frac{1}{4}(1-c_{z}+|\Omega|)$, $\omega_{2}=\frac{1}{4}(1-c_{z}-|\Omega|$, $\omega_{3}=\frac{1}{4}(1+c_{z}+|\Gamma|)$, and $\omega_{4}=\frac{1}{4}(1+c_{z}-|\Gamma|)$. Notice that the condition $S[\rho_{A}(t)]=S[\rho_{B}(t)]$ satisfied, we have the unique value of classical correlation irrespective of whether performing measurement on subsystem $A$ or $B$. In order to compute the quantum discord for Eq~(\ref{e8}), we propose the complete set of von Neumann measurement of subsystem $B$, $\{\Pi_{B}^{(i)}|i=1,2\}=\{|\phi^{(1)}\rangle\langle\phi^{(1)}|,|\phi^{(2)}\rangle\langle\phi^{(2)}|\}$, with $|\phi^{(1)}\rangle=\cos\theta|0\rangle+e^{i\varphi}\sin\theta|1\rangle$ and $|\phi^{(2)}\rangle=e^{-i\varphi}\sin\theta|0\rangle-\cos\theta|1\rangle$, the parameters $\theta,\varphi \in[0,2\pi]$. Hence, we have the remain state of subsystem $A$ with an outcome $i$,
\begin{equation}
\rho_{A}^{(i)}=\left(\begin{array}{cc}
\frac{1}{2}[1-(-1)^{i}c_{z}\cos(2\theta)] & \frac{(-1)^{i+1}}{4}(e^{-i\varphi}\Omega+e^{i\varphi}\Gamma)\sin(2\theta)\\
\frac{(-1)^{i+1}}{4}(e^{i\varphi}\Omega^{*}+e^{-i\varphi}\Gamma^{*})\sin(2\theta) & \frac{1}{2}[1+(-1)^{i}c_{z}\cos(2\theta)]
\end{array}\right)\\
\end{equation}
and the respective probability $p_{i}=\frac{1}{2}$. Consequently, we obtain the classical correlation of Eq.~(\ref{e8})
\begin{equation}
C[\rho_{AB}(t)]=\frac{1+\vartheta}{2}\log_{2}(1+\vartheta)+\frac{1-\vartheta}{2}\log_{2}(1-\vartheta),
\end{equation}
where $\vartheta=\max\bigg\{|c_{z}|,\frac{|\Omega|+|\Gamma|}{2}\bigg\}$. Finally, the quantum discord is given by
\begin{eqnarray}\label{e10}
Q[\rho_{AB}(t)] & = & \frac{1}{4}[(1-c_{z}+|\Omega|)\log_{2}(1-c_{z}+|\Omega|) \nonumber \\
 &  & +(1-c_{z}-|\Omega|)\log_{2}(1-c_{z}-|\Omega|)\nonumber \\
 &  & +(1+c_{z}-|\Gamma|)\log_{2}(1+c_{z}-|\Gamma|)\nonumber \\
 &  & +(1+c_{z}+|\Gamma|)\log_{2}(1+c_{z}+|\Gamma|)]-C[\rho_{AB}(t)] \nonumber \\
\end{eqnarray}

Entanglement is well known as a kind of quantum correlation, which is different from quantum discord.
In order to investigate the relation between quantum discord and entanglement, we choose the entanglement
of formation ($E$) for quantifying the amount of entanglement of two-qubit. The $E$ is a monotonically
increasing function of concurrence ($C^{\prime}$) \cite{Wootters1},
\begin{equation}\label{e11}
E=-f(C^{\prime})\log_{2}f(C^{\prime})-[1-f(C^{\prime})]\log_{2}[1-f(C^{\prime})],
\end{equation}
with $f(C^{\prime})=(1+\sqrt{1-C^{\prime2}})/2$. The concurrence
$C^{\prime}$ is defined as
$C^{\prime}=\max\{0,2\xi_{max}-Tr[\sqrt{\rho_{AB}\tilde{\rho}_{AB}}]\}$,
where
$\tilde{\rho}_{AB}=\sigma_{A}^{y}\otimes\sigma_{B}^{y}\rho_{AB}\sigma_{A}^{y}\otimes\sigma_{B}^{y}$,
$\xi_{max}$ is the maximum eigenvalue of
$\sqrt{\rho_{AB}\tilde{\rho}_{AB}}$. One can directly calculate concurrence for the state given by Eq.~(\ref{e8}), $C^{\prime}=\max\bigg\{0,\frac{|\Gamma|+c_{z}-1}{2},\frac{|\Omega|-c_{z}-1}{2}\bigg\}$.

\subsection{evolution from pure state}
To investigated time evolution of quantum correlations of the two-qubit initially prepared in the pure state, we set the parameters $c_{x}=1$, and $-c_{y}=c_{z}=1$, then the initial state becomes Bell state $\frac{1}{\sqrt{2}}(|00\rangle+|11\rangle)$. Substituting these values into Eq.~(\ref{e10}), and noticing that $\vartheta=1$, i.e., classical correlation always equals to $1$, finally, we obtain the quantum discord for this case,
$Q=\frac{1-|F_{14}(t)|}{2}\log_{2}(1-|F_{14}(t)|)+\frac{1+|F_{14}(t)|}{2}\log_{2}(1+|F_{14}(t)|)$.
Obviously, $Q$ is a monotonically increasing function of variable $|F_{14}(t)|$, gets its maximum ($Q=1$) with $|F_{14}(t)|=1$ and minimum ($Q=0$) with $|F_{14}(t)|=0$. In a previous work, Yuan \textit{et al}. revealed $|F(t)|$ decay to zero in short time evolution when $\lambda$ approaches the critical point $\lambda_c=1$ \cite{Yuan1} in the weak coupling regime ($g\ll1$). Thus one can expect that $Q$ just vanishes at the critical point. To evaluate the entanglement, we need to calculate the concurrence and it is $C^{\prime}=|F_{14}(t)|$ for this case. From above discussion, both quantum discord and entanglement are the monotonically increasing function of variable $|F_{14}(t)|$. Thus, entanglement behaves in a similar way as quantum discord does. In Fig.~\ref{FIG1}, we plot the time evolution of quantum discord and entanglement for Ising chain ($\gamma=1$), and other parameters $g=0.05, \delta=0, D=0, N=600$ at different points $\lambda=0.85, 1.00, 1.15$. Quantum discord is always less than entanglement in the process of evolution. At point $\lambda=1$, quantum discord and entanglement jointly decay to zero at the same time.

Now, we consider the effects of DM interaction on quantum correlation. First, let us check the dynamical property of correlation at critical point. One may recall the approximation of $|F_{14}(t)|$ given in Ref.\cite{Cheng1}. Following the similar procedure, one can introduce a cutoff number $K_{c}$ and define the partial product for $|F_{14}(t)|$, $|F_{14}(t)|_{c}=\prod_{k>0}^{K_{c}}F_{k}(t)\geq|F_{14}(t)|$. The partial product can be written as $S(t)=\ln|F_{c}(t)|=-\sum_{k=\frac{1}{2}}^{K_{c}}\ln|F_{k}(t)|$. One has $\Lambda_{k}^{(\lambda)}\approx|\lambda-1|+4D\sin\bigg(\frac{2\pi k}{N}\bigg)$ and $\Lambda_{k}^{(\lambda_{\mu})}\approx|\lambda_{\mu}-1|+4\sin\bigg(\frac{2\pi k}{N}\bigg)$, $(\mu=1, 4)$ for small $k$ and large $N$. Thus one has $\sin(2\Theta_{k}^{(\lambda_{\mu})})=\frac{\mp2\pi\gamma gk}{|(\lambda_{\mu}-1)(\lambda-1)|}$, $(\mu=1, 4)$ and $\sin(\Theta_{k}^{(\lambda_{1})}-\Theta_{k}^{(\lambda_{4})})\approx\frac{-2\pi\gamma gk}{|(\lambda_{1}-1)(\lambda_{4}-1)|}$. Then, omitting the terms related to the sum of $k^{4}/N^{4}$, one obtain the approximation for the partial product $S(t) \approx -\frac{1}{2}\frac{4\pi\gamma^{2}g^{2}}{N^{2}(\lambda-1)^{2}}\sum_{k=\frac{1}{2}}^{K_{c}}k^{2}
\bigg\{\frac{\sin^{2}(\Lambda_{k}^{(\lambda_{4})}t)}{(\lambda_{4}-1)^{2}}+\frac{\sin^{2}(\Lambda_{k}^{(\lambda_{1})}t)}{(\lambda_{1}-1)^{2}}+
2\frac{\sin(\Lambda_{k}^{(\lambda_{1})}t)\sin(\Lambda_{k}^{(\lambda_{4})}t)}{|(\lambda_{1}-1)(\lambda_{4}-1)|}\cos(4gt)\bigg\}$. Finally, when $\lambda\rightarrow1$, in the short time one has $|F_{14}(t)|\approx e^{-(\tau_{1}+\tau_{2})t^{2}}$ with $\tau_{1}=32\pi^{2}\gamma^{2}g^{2}/(\lambda-1)^{2}\sum_{k}^{K_{c}}k^{2}/N^{2}$ and $\tau_{2}=256\pi^{3}\gamma^{2}gD/(\lambda-1)^{2}\sum_{k}^{K_{c}}k^{3}/N^{3}$. Consequently, decay of quantum discord and entanglement may be enhanced by introducing DM interaction. The numerical calculation plotted in Fig.~\ref{FIG2} shows the effects of DM interaction, which is consistent with theoretical
computing. It is found that the smaller values of $\gamma$, the stronger the effect of DM interaction on the decay of the decoherence factor \cite{Cheng1}. It implies that the effect of DM interaction on quantum correlation may be controlled by the parameter $\gamma$.

\begin{figure}
  \includegraphics[width=6cm]{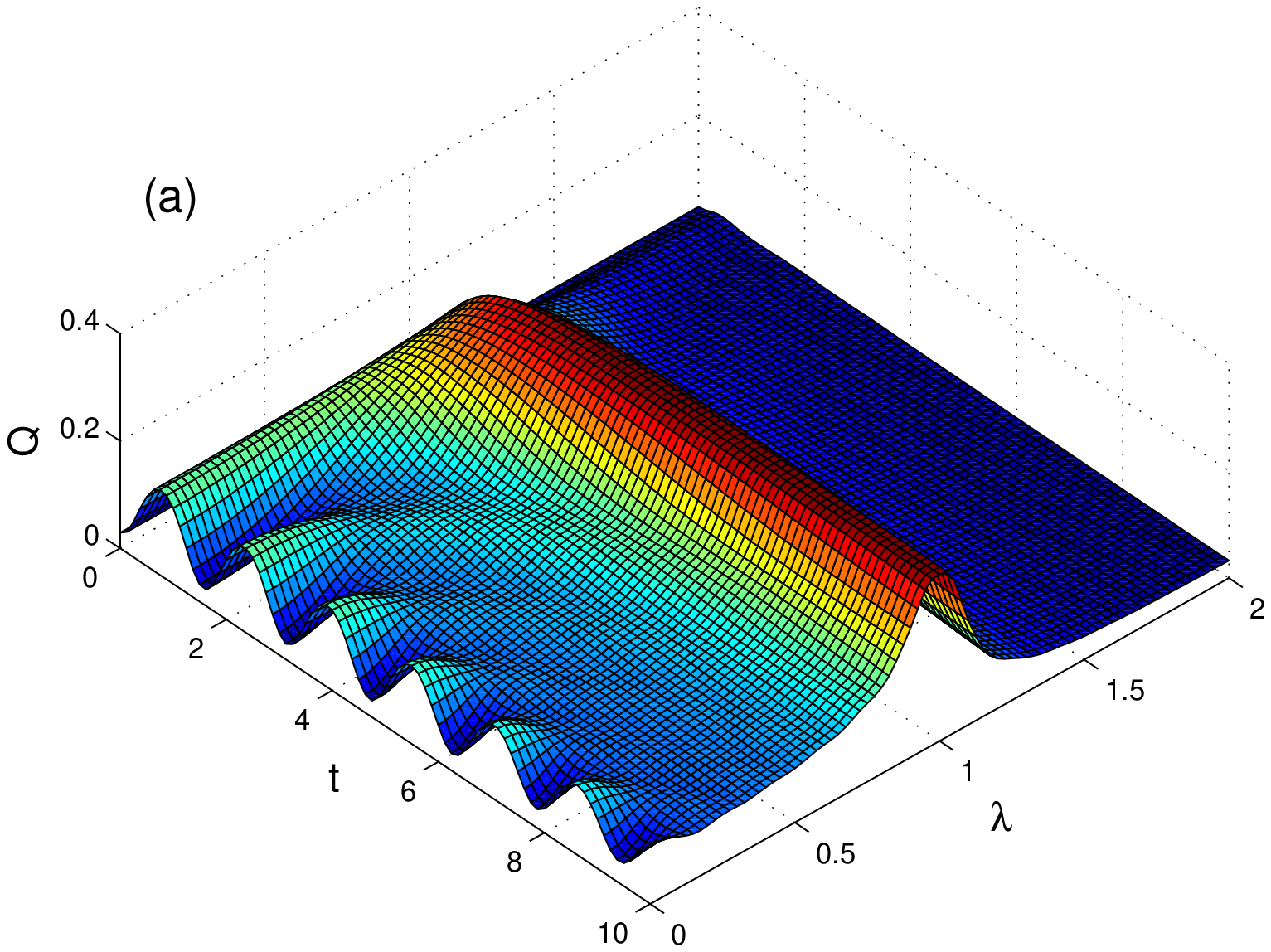}\\
  \includegraphics[width=6cm]{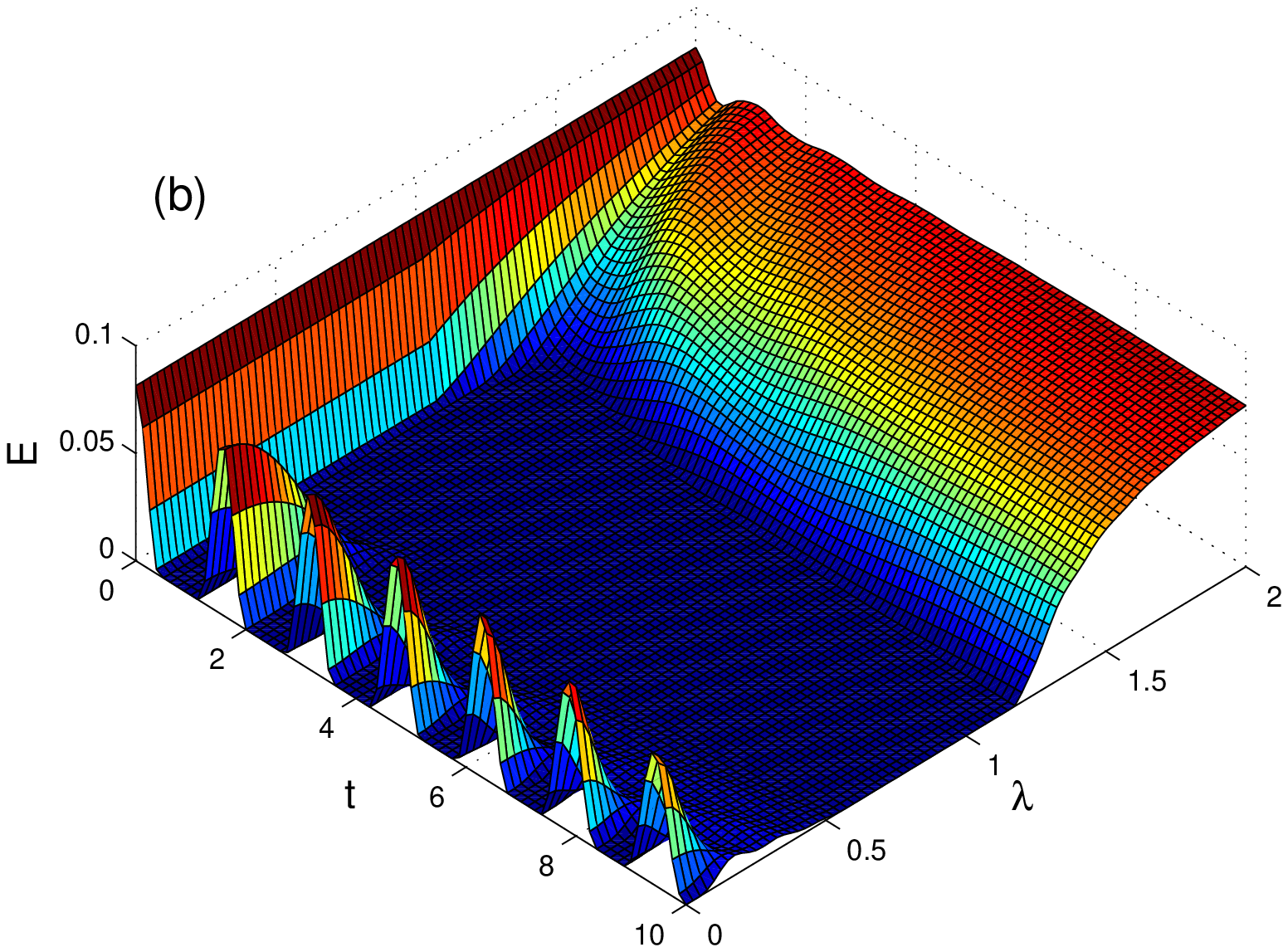}\\
  \caption{(a) Quantum discord and (b) entanglement as a function of time $t$ and $\lambda$. Other parameters are set as $\gamma=1, g=0.05, \delta=0, D=0$, and $N=600$.}\label{FIG3}
\end{figure}

\subsection{evolution from mixed state}

In this subsection, we study the time evolution of correlations when two-qubit initially prepared in the mixed state. We choose the parameters $c_{x}=1, -c_{y}=c_{z}=0.2$, thus the initial state becomes a mixed state $\rho_{AB}(0)=0.6|\Phi^{+}\rangle\langle\Phi^{+}|+0.4|\Psi^{+}\rangle\langle\Psi^{+}|$, where $|\Phi^{+}\rangle=\frac{1}{\sqrt{2}}(|00\rangle+|11\rangle)$ and $|\Psi^{+}\rangle=\frac{1}{\sqrt{2}}(|01\rangle+|10\rangle)$, and it is easy to check that Eq.~(\ref{e7}) is a family of Bell-diagonal state for $c_{x}=1$, $c_{y}=-c_{z}, c_{z}\in[0, 1]$. To illustrate the dynamical properties of quantum discord and entanglement, we carry out the numerical calculation from the exact expression (\ref{e10}) and (\ref{e11}). In Fig.~\ref{FIG3}, quantum discord and entanglement are plotted as a function of time $t$ and $\lambda$ with the parameters $g=0.05, \delta=0, \lambda=1, D=0, N=600$. It is observed that quantum discord gets maximized at critical point but entanglement rapidly decays to zero and vanishes in the area where quantum discord has nonzero values. This phenomenon is also different from the former case of pure state. It implies environment may enhance quantum discord of two-qubit.

\begin{figure}
  \includegraphics[width=8cm]{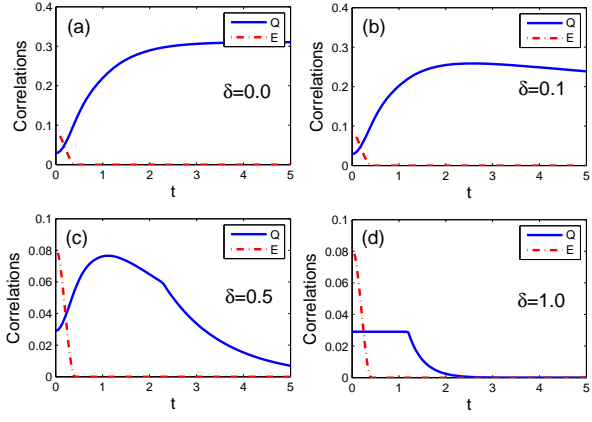}\\
  \caption{Quantum discord (solid line) and entanglement (dash-dot line) as a function of time $t$ and different $\delta$. Other parameters are set as $\gamma=1, g=0.05, D=0$, and $N=600$, (a) $\delta=0.0$, (b) $\delta=0.1$, (c) $\delta=0.5$, (d) $\delta=1.0$.}\label{FIG5}
\end{figure}

\begin{figure}
  \includegraphics[width=8cm]{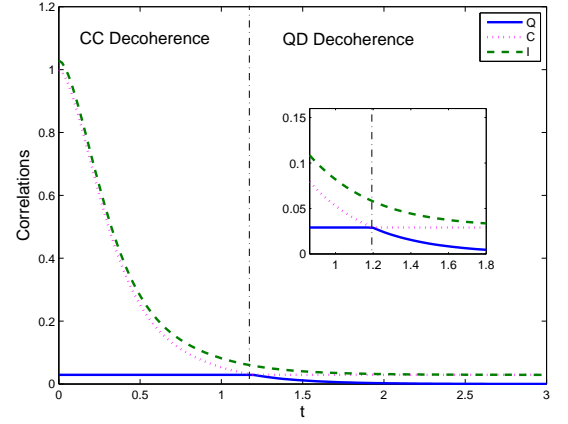}\\
  \caption{Quantum discord (solid line), classical correlation (dot line), and total correlation (dash line) as a function of time $t$. Other parameters are set as $\gamma=1, \lambda=1, g=0.05, \delta=1.0, D=0$, and $N=600$.}\label{FIG7}
\end{figure}

\begin{figure}
  \includegraphics[width=8cm]{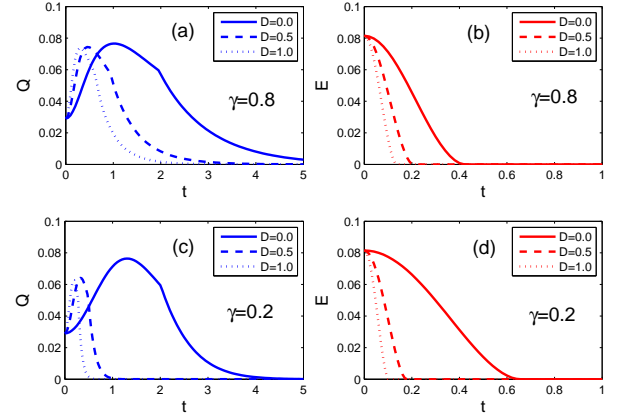}\\
  \caption{(a) and (c) plots of quantum discord as a function of time $t$ under different $D$, and (b) and (d) plots of entanglement as a function of time $t$ under different $D$. Other parameters are set to $g=0.05, \delta=0.5$, and$N=600$, (a) and (b) $\gamma=0.8$, (c) and (d) $\gamma=0.2$.}\label{FIG6}
\end{figure}

Let us turn to study the effects of the parameters $\delta$ on the quantum correlation. For the case of $\delta=0$, the decoherence factor $|F_{23}(t)|=1$, otherwise one has $|F_{23}(t)|\neq1$, and particularly, $|F_{23}(t)|=|F_{14}(t)|$ when $\delta=1$. In Fig.~\ref{FIG5}(a)-(d), we plot the  quantum discord and entanglement against time $t$ with $\delta$ equals to $0.0, 0.1, 0.5, 1.0$, respectively. It is observed that the decay of  quantum discord can be enhanced by increasing $\delta$, however, entanglement seems to be no sensitive to the change of $\delta$. When $\delta=1$, it means one qubit interacts with the surrounding chain and the other is free from the environment. Recalling that $|F_{14}(t)|$ decreases to zero in a very short time under some reasonable condition, and so does $|F_{23}(t)|$. Thus, the nondiagonal elements in expression (\ref{e8}) vanish all at once, that is to say the coherence of two-qubit quickly disappears. As plotted in the Fig.~\ref{FIG5}(d), quantum discord remains constant for an interval of time, and then decays to zero. This situation seems to be sudden transition between classical and quantum decoherence, which has been detail discussed in the Ref.~\cite{Mazzola1}. Figure.~\ref{FIG7} shows that the situation is indeed the transition. In the left side of dash-dot line in the plot, one can observe that classical correlation decreases and quantum discord maintain constant. Whereas, in the other side of the line, classical correlation do not change with time and quantum discord start decreasing.

To end this subsection, let us focus on how the DM interaction affects the quantum correlation. In Fig.~\ref{FIG6}, we plot the time evolution of quantum discord and entanglement when two-qubit coupled to a general $XY$ chain under different $D$. It is observed that the decay of quantum discord can be slightly enhanced by decreasing values of $\gamma$ for case of $D=0$. However, we find that the decay of entanglement can be slightly increased by increasing $\gamma$, which is opposed from that of quantum discord. Nevertheless, entanglement quickly vanishes in a much shorter time than quantum discord does. Fig.~\ref{FIG6} also shows that the effects of DM interaction on quantum discord is more remarkable than its on entanglement.

\section {Conclusions}
In summary, we have studied the time evolution of quantum discord and entanglement for two-qubit coupled to a spin chain with DM term. We evaluated the  quantum discord and entanglement for two-qubit to be prepared in a class of $X-$structure state. We have separately studied the two-qubit evolutes from pure state and mixed state. The difference of quantum discord and entanglement becomes drastic in the case of mixed state. In the case of pure state, it is found that quantum correlation rapidly decays to zero at critical point where environment has a QPT. Moreover, the DM interaction can enhance the decay of quantum correlation for the case. Interestingly, quantum discord may get maximized in the case of mixed state when environment at the critical point while entanglement vanishes when environment approaches the critical point. We also considered that qubit coupled to the environment with different coupling strength from each other's, which is controlled by the parameter $\delta$. It has been shown that the coherence of two-qubit rapidly vanishes when single qubit interacts with the environment. Besides, we have observed sudden transition between classical and quantum decoherence for the case. The effect of DM interaction in this case is not only enhancing the decay of quantum discord and entanglement, but also enhancing the increasing of quantum discord. Furthermore, the effect of DM interaction on quantum discord can be strengthened by the anisotropy parameter. However, the DM interaction has slight effect on entanglement with respect to that of quantum discord.

\begin{acknowledgments}
This work was partially supported by the NSF of China (Grant No.
11075101), and Shanghai Research Foundation (Grant No. 07dz22020).
\end{acknowledgments}

\end{document}